\documentclass[aps,prb,twocolumn,amssymb,a4paper,longbibliography]{revtex4-1}

\usepackage{graphicx}
\usepackage{amsmath}
\usepackage{bm}
\usepackage{hyperref}
\usepackage{color}

\def\vR{{\bf R}}
\def\vJ{{\bf J}}
\def\vd{{\bm{\delta}}}
\def\vk{{\bf k}}
\def\vkF{{\bf k}_{\mathrm{F}}}
\def\vv{{\bf v}}
\def\vvF{{\bf v}_{\mathrm{F}}}
\def\vF{v_{\mathrm{F}}}

\def\NF{N_{\mathrm{F}}}
\def\FS{_{\mathrm{FS}}}
\def\ua{{\hat{\bf a}}}
\def\ub{{\hat{\bf b}}}
\def\barvF{\bar{v}_{\mathrm{F}}}

\begin{document}

%\title{Breaking time-reversal symmetry in mesoscopic superconducting grains}
\title{Breaking time-reversal and translational symmetry at edges of $d$-wave superconductors:
microscopic theory and comparison with quasiclassical theory}
\author{N. Wall Wennerdal}
\author{A. Ask}
\author{P. Holmvall}
\author{T. L\"ofwander}
\author{M. Fogelstr\"om}
\affiliation{Department of Microtechnology and Nanoscience - MC2,
	Chalmers University of Technology, SE-412 96 G\"oteborg, Sweden}

\date{\today}

\begin{abstract}
We report results of a microscopic calculation of a second-order phase transition into a state breaking time-reversal and translational invariance along pair-breaking edges of $d$-wave superconductors. By solving a tight-binding model through exact diagonalization with the Bogoliubov-de~Gennes method, we find that such a state with current loops having a diameter of a few coherence lengths is energetically favorable below $T^*$ between 10\%-20\% of $T_{\mathrm{c}}$ of bulk superconductivity, depending on model parameters. This extends our previous studies of such a phase crystal within the quasiclassical theory of superconductivity, and shows that the instability is not qualitatively different when including a more realistic band structure and the fast oscillations on the scale of the Fermi wavelength. Effects of size quantization and Friedel oscillations are not detrimental. We also report on a comparison with quasiclassical theory with the Fermi surfaces extracted from the tight-binding models used in the microscopic calculation. There are quantitative differences in for instance the value of $T^*$ between the different models, but we can explain the predicted transition temperature within each model as due to the different spectral weights of zero-energy Andreev bound states and the resulting gain in free energy by breaking time-reversal and translational invariance below $T^*$.
\end{abstract}

\maketitle

\section{Introduction}

High-temperature superconductors have been shown to have a superconducting order parameter of $d$-wave symmetry.\cite{vanHarlingen:1995,Tsuei:2000} Although this is well established, a lot of research is still ongoing to explain the full phase diagram, and the mechanism of superconductivity is still controversial.\cite{Scalapino:2012,Fradkin:2015} In addition, there are remaining interesting, and somewhat controversial, questions on the properties of the $d$-wave superconducting phase, in particular in devices where surfaces and interfaces play an important role.\cite{Hilgenkamp:2002,Gustafsson:2012} Surfaces and interfaces with a normal not exactly aligned with a crystallographic axis are pair breaking, with associated formation of zero-energy Andreev bound states.\cite{Hu:1994} These states play an important role in determining device physics. They show up in tunneling experiments as zero-bias conductance peaks,\cite{Kashiwaya:2000} and influence the current-phase relation of Josephson junctions.\cite{Lofwander:2001,Golubov:2004}

From a fundamental physics point of view, it is interesting that the flat band of zero-energy Andreev bound states can be related to bulk topology,\cite{Sato:2011} but at the same time may lead to instabilities where additional symmetries are broken.\cite{Sigrist:1998} Early on it was suggested that time-reversal symmetry may be broken by forming a subdominant component of the order parameter,\cite{Matsumoto:1995,Sigrist:1996,Fogelstrom:1997,Tanuma:1998} thereby shifting the Andreev states and splitting the zero-bias peak as sometimes seen experimentally.\cite{Covington:1997,dagan:2001} Another suggestion is magnetic ordering at the surface causing a spin split.\cite{Honerkamp:2000,Potter:2014} Experimentally, breaking of time-reversal symmetry remains controversial.\cite{Tsuei:2000,Carmi:2000,Neils:2002,Kirtley:2006,Saadaoui:2011,Gustafsson:2012} The zero-bias conductance peak does not always split at low temperature, and the associated currents and magnetic fields are so far unobserved or small. These works may also become relevant to the ongoing research into Sr$_2$RuO$_4$, where recent experiments\cite{Pustogow:2019,Ishida:2020} point towards a singlet superconductor with unconventional orbital symmetry.\cite{Ramires:2019,Kivelson:2020}

Recently, we have reported on a different scenario of shifting the Andreev bound states and lowering the free energy in a more complicated manner, where both time-reversal and translational invariance along the surface are broken.\cite{Hakansson:2015,Holmvall:2018a,Holmvall:2018b,Holmvall:2019,Holmvall:2020} In a second order phase transition, spontaneous current loops become energetically favorable at a temperature $T^*$ up to 20\% of $T_{\mathrm{c}}$ of bulk superconductivity. The inhomogeneous broadening along the surface may explain that the zero-bias conductance peak is not necessarily split in a tunneling experiment, but instead acquires a temperature-independent width for $T<T^*$. In addition, since neighboring current loops have opposite circulation, the magnetic fields tend to cancel when averaged over one period of the current pattern, which is about $10\xi_0$, where $\xi_0$ is the superconducting coherence length.\cite{Hakansson:2015} This unexpected symmetry-broken state does not involve any subdominant pairing channel or any other mean-field order than the superconducting order parameter with pure $d$-wave symmetry. The free energy is lowered through Doppler shifts of Andreev states. The superflow causing the Doppler shift is associated with the order-parameter phase having an oscillatory behavior along the edge, resembling a crystal pattern that may be called a phase crystal.\cite{Holmvall:2020}
The phase crystal has a higher $T^*$ than the state with translational invariant superflow.\cite{Higashitani:1997,Fauchere:1999,Barash:2000,Lofwander:2000,Suzuki:2014}

The previous work\cite{Hakansson:2015,Holmvall:2018a,Holmvall:2018b,Holmvall:2019,Holmvall:2020}
and predictions were carried out using the quasiclassical theory of superconductivity.\cite{Serene:1983}
The quasiclassical approximation involves integrating out effects relevant at the atomic scale.
This requires a good separation between the atomic scale and the relevant superconducting scale, i.e. the superconducting coherence length.
Typical high-temperature superconductors have very short coherence lengths, and the validity of the quasiclassical approximation can be questioned.
A first comparison of results from a tight-binding Bogoliubov-deGennes theory with those of quasiclassical theory\cite{BlackSchaffer:2013} were made analysing the unconventional charging effects in small superconducting $\mathrm{YBCO}$ single-electron tunneling devices\cite{Gustafsson:2012}.  
In this paper we explore the phase crystal at edges of $d$-wave superconductors within a tight-binding model.
We are interested in the effects of including the atomic scale oscillations,
as well as the effects of the more realistic band structure and Fermi surface taken into account in the tight-binding model.
So far\cite{Hakansson:2015,Holmvall:2018a,Holmvall:2018b,Holmvall:2019,Holmvall:2020} only a circular Fermi surface
was used in the quasiclassical calculations. We will in this paper also extract the relevant Fermi surfaces predicted by the tight-binding
models and compare with quasiclassical theory.

The paper is structured as follows. In Section~\ref{TBmodel} we introduce the real-space tight-binding model and how it is solved using the Bogoliubov-de~Gennes exact diagonalization method. This section also includes a reciprocal space calculation to characterize the parameter spaces, set key model parameters, and extract the relevant Fermi surfaces. In Section~\ref{sec:results} we present results obtained within the tight-binding model for the transition into the phase crystal. In Section~\ref{sec:comparison} we introduce a comparison between the tight-binding model and the quasiclassical formulation with the extracted Fermi surfaces, thereby highlighting the qualitative similarities but quantitative differences between the two approaches. Finally in Section~\ref{sec:summary} we summarize. A few calculations have been collected in the Appendix.

\section{Tight-binding description of a d-wave superconductor}
\label{TBmodel}

\subsection{Real space formulation}
\begin{figure}
	\includegraphics[width=0.9\columnwidth]{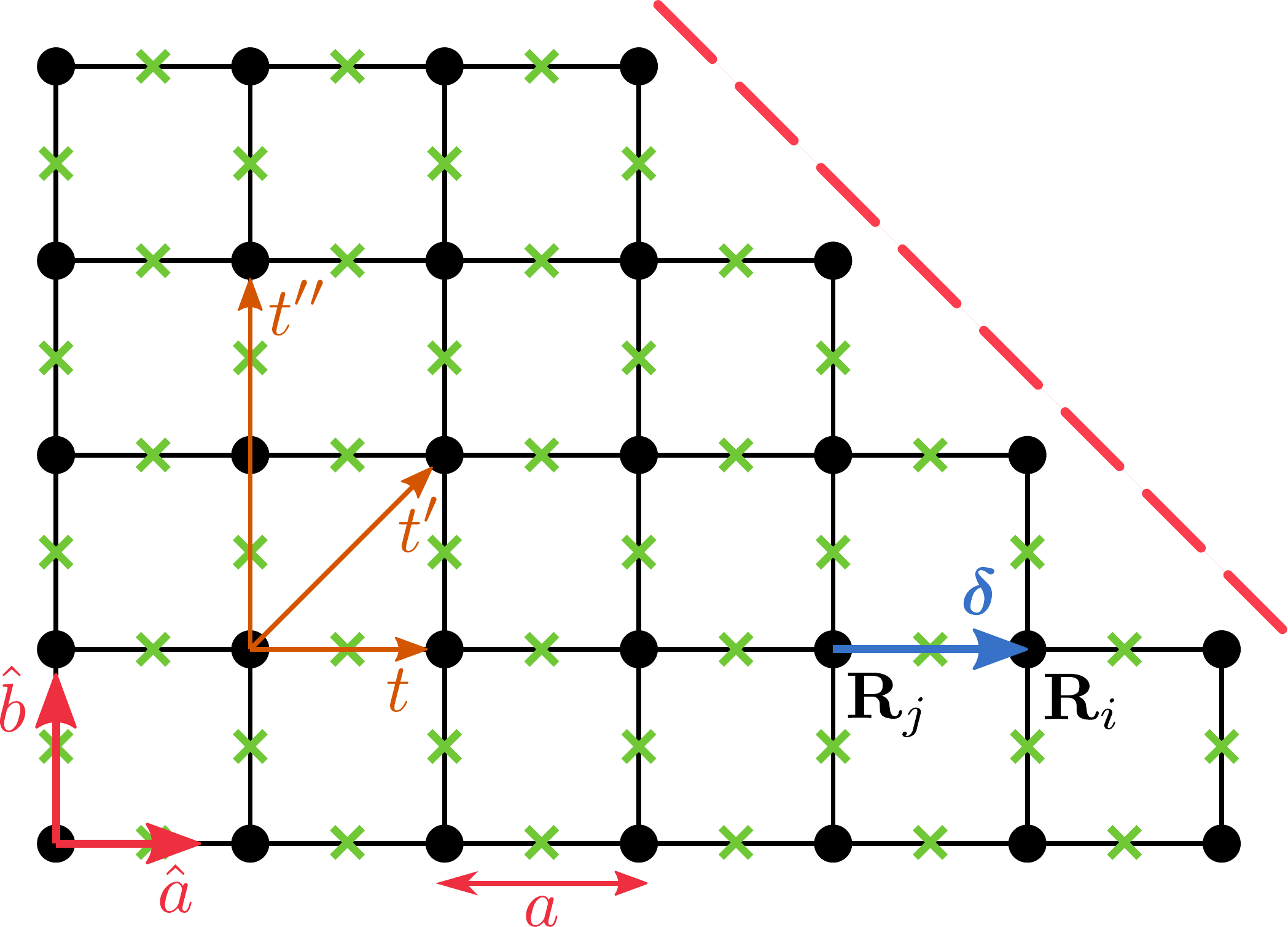}
	\caption{Illustration of the tight-binding model. The lattice sites are indicated by filled black circles and have coordinates $\vR_j$. The hopping integrals included are nearest $t$, next-nearest $t'$, and next-next-nearest $t''$ neighbors. The lattice constant is $a$ and we assume that it is equal along both crystallographic directions $\ua$ and $\ub$. The link order parameter $\Delta_{ij}$ is defined along links such as $\vd=\vR_i-\vR_j$ and we assign their values to the midpoints between sites as indicated by green crosses. The dashed line indicates a pair-breaking [110]-edge, where zero-energy states are formed and where the circulating currents appear below $T<T^*$.}
	\label{lattice}
\end{figure}

The normal state of the material is described by a single band, where the bandstructure is determined by a few hopping integrals $t_{ij}$.
The single-particle Hamiltonian reads
\begin{equation}
	\hat{ \mathcal{H}}^{(e)}=\sum_{i,j,\sigma}\hat c^{\dagger}_{i\sigma} \mathcal{H}^{(e)}_{ij} \hat c_{j\sigma} =
	\sum_{i,j,\sigma}\hat c^{\dagger}_{i\sigma} (t_{ij}-\mu\delta_{ij})\hat c_{j\sigma},
	\label{H_e}
\end{equation}
where $t_{ij}$ includes nearest ($t$), next-nearest ($t'$), and next-next-nearest neighbor ($t''$) hopping parameters,
see Fig.~\ref{lattice}. The indices $i$ and $j$ enumerate the lattice sites, $\sigma$ labels spin, and $\delta_{ij}$ is the Kronecker delta-function.
The operator $\hat c_{j\sigma}$ annihilates an electron with spin $\sigma$ on site $j$.
The chemical potential $\mu$ is set by the doping level, but is here treated as a parameter of the model.
We take the hopping parameters from literature\cite{Radtke:1994,Berthod:2017} where they have been extracted either from relevant experimental data or from density functional theory.

%%%
%%%

The Hamiltonian describing weak coupling $d$-wave superconductivity on a lattice is
\begin{equation}
	\hat H_{d} = \sum_{\langle i,j\rangle} \Delta_{ij}\hat c_{i\uparrow}^{\dagger} \hat c_{j\downarrow}^{\dagger} + h.c.
\end{equation}
where the mean field order parameter lives on nearest neighbor links
\begin{equation}
	\Delta_{ij} = \frac{V}{2}\langle \hat c_{i\uparrow} \hat c_{j\downarrow} - \hat c_{i\downarrow} \hat c_{j\uparrow} \rangle,
\end{equation}
where $V$ is the coupling constant.
In the bulk, $\Delta_{ij}$ are positive for links along the crystallographic $\ua$-axis, $\vR_i-\vR_j=\pm a\ua$,
and negative for links along the $\ub$-axis, $\vR_i-\vR_j=\pm a\ub$, where $a$ is the lattice constant.
We note that depending on the parameters of the model (see next subsection), an extended $s$-wave order
may be stable instead of the $d$-wave, in which case the link order parameter $\Delta_{ij}$ is positive in both crystallographic directions.
In this paper, we will focus on the part of parameter space where $d$-wave order is stable in the bulk.

After performing a Bogoliubov rotation, the Bogoliubov-de~Gennes (BdG) equation for a superconductor is obtained
\begin{equation}
	\left(\begin{array}{cc}
			{\mathcal{H}}^{(e)}_{ij} & \Delta_{ij}                \\
			\Delta^*_{ij}            & -{\mathcal{H}}^{(e)*}_{ij}\end{array}\right)
	\left(\begin{array}{c} u^{(n)}_j \\v^{(n)}_j \end{array}\right)
	= E_n
	\left(\begin{array}{c} u^{(n)}_i \\v^{(n)}_i \end{array}\right),
	\label{BdGequation}
\end{equation}
where  $E_n$ is the eigenvalue and $\left(u^{(n)}_i,\,\, v^{(n)}_i\right)^T$ is the corresponding eigenfunction.
Since we are considering a singlet superconductor, we are here suppressing the spin indices but consider both
positive and negative energy eigenvalues, thereby avoiding double counting.
Note that if the $u$-amplitudes are spin up, then the $v$-amplitudes are spin down, see e.g. Ref.~\onlinecite{Zhu:book}.
The most straightforward strategy is to resort to direct diagonalization and obtain all eigenvalues and eigenvectors.
The only complication is that the link order parameter must be computed self-consistently through the gap equation
\begin{equation}
	\Delta_{ij}=\frac{V}{4}\sum_n\,\bigg(u^{(n)}_i v^{(n)*}_j+v^{(n)*}_i u^{(n)}_j\bigg)\tanh \left (\frac{E_n}{2T} \right ),
	\label{Delta_link}
\end{equation}
where $T$ is the temperature. Note that in this paper we set the reduced Planck constant $\hbar$, the Boltzmann constant $k_{\mathrm{B}}$, and the electron charge $e$, all to unity.

Once self-consistency of $\Delta_{ij}$ has been achieved we may compute observables, such as currents and local density of states.
The local current density is in the tight-binding model defined on links coupling nodes through the hopping integrals $t_{ij}$ in Eq.~(\ref{H_e}).
The current from site $j$ to site $i$ is computed through the formula
\begin{equation}
	J_{ij}=-4 \sum_n \mbox{Im}\bigg\lbrack t_{ij}  u^{(n)*}_i u^{(n)}_j f(E_n) -t^{*}_{ij} v^{(n)*}_i v^{(n)}_j (1-f(E_n))\bigg\rbrack
	\label{current}
\end{equation}
where $f(E_n)$ is the Fermi-Dirac distribution function. We use as an extra convergence test that
the currents flowing into and out of all nodes in the grain are conserved.

The local density of states at position $\vR_j$ is computed as
\begin{equation}
	{\cal N}_{j}(E) = -\frac{1}{\pi}\Im\sum_n \left[
		\frac{\vert\langle u^{(n)}_{j}\vert u^{(n)}_{j}\rangle\vert^2}{E-E_n+i \eta} +
		\frac{\vert\langle v^{(n)}_{j}\vert v^{(n)}_{j}\rangle\vert^2}{E+E_n+i \eta} \right],
	\label{LDos}
\end{equation}
where $\eta>0$ is a small imaginary part of the energy.
The total density of states of the grain is obtained by summation over sites.

Once we have the full eigenvalue spectrum we can also study the thermodynamic properties. The free energy is given as \cite{Kosztin1998}
\begin{eqnarray}
	\Omega&=&E_g-T\mathcal{S}\nonumber\\
	&=&-T\sum_n\ln\bigg\lbrack 2 \cosh\left(\frac{E_n}{2T}\right)\bigg\rbrack
	+\sum_{i,j}\frac{|\Delta_{ij}|^2}{2 V},
	\label{free-energy}
\end{eqnarray}
while the entropy has the well-known form
\begin{equation}
	\mathcal{S}=-2\sum_n\bigg\lbrack f(E_n) \ln f(E_n)+(1-f(E_n)) \ln(1-f(E_n))\bigg\rbrack.
	\label{entropy}
\end{equation}

We also use the convergence of the free energy as a check for our solutions.
In principle, Eq. (\ref{free-energy})  contains an additional term,
\begin{equation}
	\sum_{n,i} E_n(|u^{(n)}_i|^2-|v^{(n)}_i|^2),
\end{equation}
that stems from the internal energy $E_g$. This term gives, when non-zero, the
same contribution in both the normal and in the superconducting states and is omitted, since we will present results for the free energy difference between the superconducting and normal states.

\subsubsection{Notes on site versus link quantities}

In the tight-binding model it is important to keep in mind that some quantities are defined on lattice sites, while others are defined on the links.
For instance, when plotting the current it is sometimes convenient to define a vector field with arrows residing on the lattice sites.
Let us introduce the following notation for the link current from site $\vR_j$ to site $\vR_i=\vR_j+\vd$,
\begin{equation}
	J_{ij} = J_j(\vd),\,\,\, \vd = \vR_i-\vR_j,
	\label{current2}
\end{equation}
which is illustrated in Fig.~\ref{lattice} for a nearest neighbor link $\vd=a\ua$.
When we in this way single out site $\vR_j$, we see that current will flow along links
to neighbors $\vR_i$ for which the hopping integrals $t_{ij}$ in Eq.~(\ref{H_e}) are non-zero.
Positive and negative values means that current flows out of or into site $j$. Current conservation requires
\begin{equation}
	\sum_{\vd} J_{j}(\vd) = 0,\,\,\, \forall j.
	\label{eq:current_conservation}
\end{equation}
Consider now the following vector field
\begin{equation}
	\vJ(\vR_j) = \sum_{\vd} J_j(\vd) \hat{\bm\delta},
\end{equation}
where ${\hat{\bm\delta}}=\vd/|\vd|$ are unit vectors.
This vector field gives a nice overview of current flow patterns.
However, it is not fulfilling current conservation\cite{Boykin:2010} and is in a strict sense an unphysical quantity.
Nevertheless, we will for convenience present also this vector field.

In the same way, in many works\cite{Soininen:1994,Franz:1996,Martin:1998,Zhu:1999,Nagai:2017} the $d$-wave order parameter is presented as a site quantity
\begin{equation}
	\Delta_d(\vR_j) = \sum_\vd s_d(\vd) \Delta_j(\vd),
	\label{site_Delta_d}
\end{equation}
where the signature function $s_d(\vd)$ is equal to $+1$ for links along the $\ua$-axis and $-1$ for links along the $\ub$-axis,
and $\Delta_j(\vd)$ is defined in analogy with Eq.~(\ref{current2}).
A caveat is that since the coupling constant $V$ for superconductivity resides on the links, it may favor an extended s-wave order instead.
That would correspond to summing with plus signs in both directions
\begin{equation}
	\Delta_s(\vR_j) = \sum_\vd \Delta_j(\vd).
	\label{site_Delta_s}
\end{equation}
In the bulk with translational invariance, these two signatures reflect the two representations $B_1$ and $A_1$ of the crystal lattice.
In a finite size system, however, both necessarily coexist which simply reflects that there is
no translational invariance and the system is inhomogeneous across the grain, for instance at the corners.
Although the nodal quantity $\Delta_s(\vR_j)$ necessarily is non-zero at all temperatures below $T_{\mathrm{c}}$,
it is always small for the parameter space studied here.
We do not consider this a subdominant order parameter component in the sense used in continuum models,\cite{Fogelstrom:1997} where there is a second coupling constant.

In Section~\ref{sec:results} we will find spontaneous currents and a complex valued link order parameter, $\Delta_{ij}=|\Delta_{ij}|\exp\left[i\chi_{ij}\right]$,
with the superconducting phase $\chi_{ij}$ oscillating along the edge.
It would then be natural to apply a gauge transform to make the order parameter real and extract the superfluid momentum,
given by the gradient of the phase\cite{Holmvall:2018b}.
However, the gauge transform is not well defined on a finite size lattice because the number of link phases $\chi_{ij}$
is not enough to uniquely compute the node phases needed\cite{Franz:2000,Vafek:2001,Vafek:2006} to define the gauge transform.
This problem does not appear for an infinite lattice\cite{Franz:2000,Vafek:2001,Vafek:2006} and is peculiar to lattices with edges.
See Ref.~\onlinecite{Feder:thesis} for a similar problem appearing when attempting to convert between
the link order parameter and the quantities in Eqs.~(\ref{site_Delta_d})-(\ref{site_Delta_s}).
In conclusion, we will in this paper focus on the self-consistently calculated link order parameter only.
When showing results, the order parameter $\Delta_{ij}$ is assigned to the midpoint $\vR=(\vR_i+\vR_j)/2$ between nodes, as indicated in Fig.~\ref{lattice}.

\subsection{Reciprocal space formulation: analysis of parameter spaces}

Before presenting results of the real space calculation it is useful to extract material parameters for an extended system both in
the normal and in the superconducting states. In addition, we extract information about the Fermi surfaces that we will use in
Section~\ref{sec:comparison} to compare the tight-binding and quasiclassical results.

\subsubsection{Normal state}

The normal state dispersion for a single band is
\begin{eqnarray}
	\epsilon_{\bf k} &=& -2 t [ \cos(k_x a) + \cos(k_ya)] \nonumber\\
	&&- 4 t' \cos(k_xa)\cos(k_ya) \label{epsilon_k}\\
	&&- 2 t'' [ \cos(2k_xa) + \cos(2k_ya) ],\nonumber
\end{eqnarray}
where $t$, $t'$, and $t''$ are the hopping integrals from Eq.~(\ref{H_e}).
In the following we use $t>0$ as the natural unit of energy in the tight-binding model.
The chemical potential is set by $\mu$, and we introduce
\begin{equation}
	\xi_{\vk} = \epsilon_{\vk} - \mu.
\end{equation}
We can then find the Fermi surface, i.e. the set of $\vkF$ that satisfies $\xi_{\vkF} = 0$. This has to be done numerically.
The velocity is defined as
\begin{equation}
	\vv_{\vk} = \nabla_{\vk} \xi_{\vk}.
\end{equation}
The expression can straightforwardly be calculated analytically.
The Fermi velocity is then obtained as $\vv_{\vkF}$ using the calculated $\vkF$.

%%%
\begin{table}[t]
	\caption{Normal state characteristics of the two tight-binding models. All energies are measured in units of $t>0$.}
	\begin{tabular}{|c|c|c|c|c|c|c|}
		\hline\hline
		\# & $t'$   & $t''$ & $\mu$  & $\frac{\langle |\vv_{\vkF}|\rangle_{\mathrm{FS}}}{ta}$ & $\frac{1}{a\langle |\vkF|\rangle_{\mathrm{FS}}} $ \\
		\hline
		\hline
		1  & -0.25  & 0.0   & 0.0    & 2.53562                                                & 0.381869                                          \\
		\hline
		2  & -0.495 & 0.156 & -1.267 & 2.39512                                                & 0.480474                                          \\
		\hline\hline
	\end{tabular}
	\label{table:tB-models}
\end{table}
%%%

\begin{figure}
	\includegraphics[width=\columnwidth]{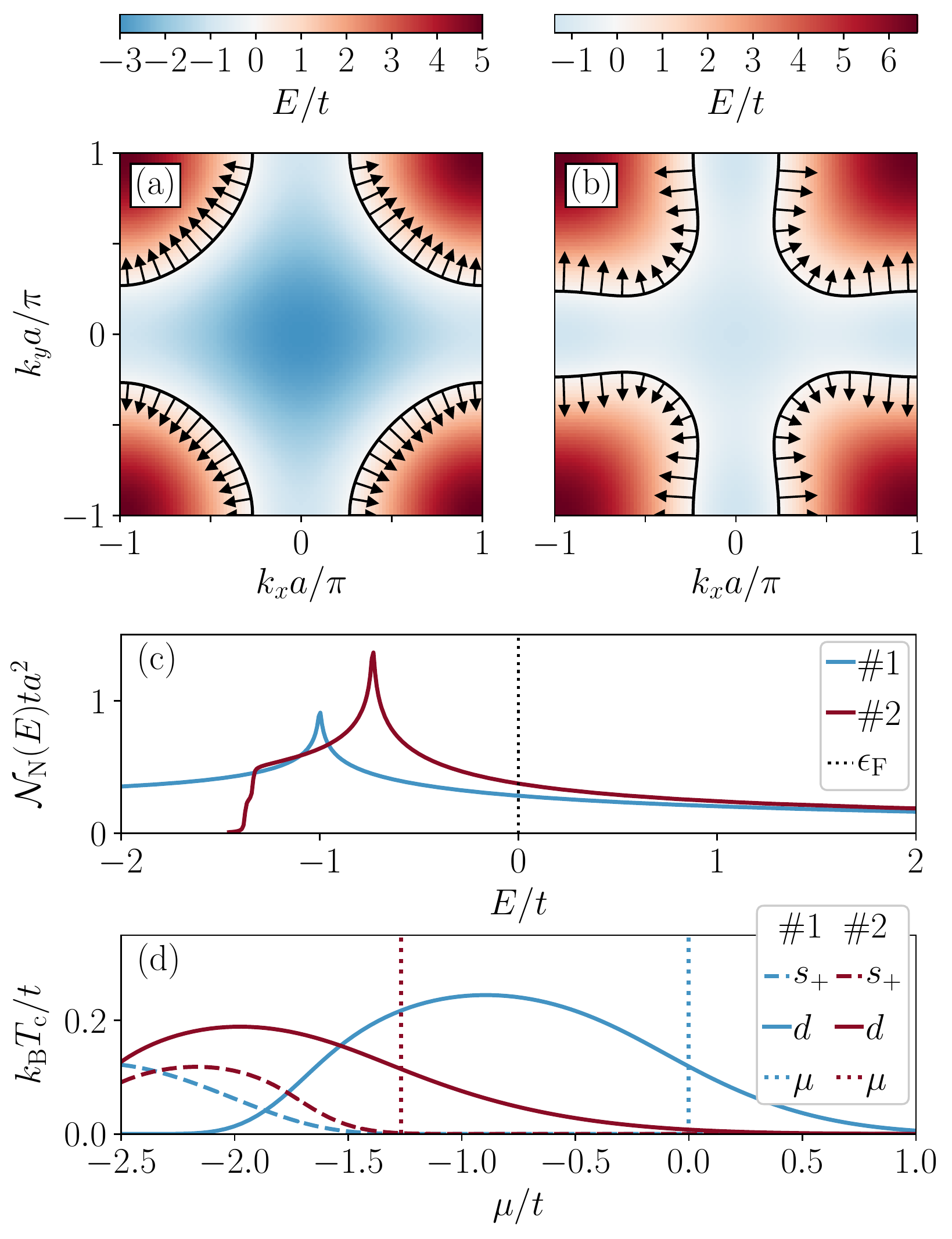}
	\caption{Bandstructure for tight-binding model \#1 (a) and model \#2 (b). The Fermi surfaces are marked with black solid lines. The arrows show how the Fermi velocity varies around the Fermi surfaces. (c) The resulting density of states. (d) Critical temperature $T_{\mathrm{c}}$ for bulk superconductivity for both $d$-wave and extended $s$-wave superconductivity.
		The light blue lines are for model \#1 and the dark red lines are for model \#2.
		The position of the Fermi energy (dotted lines) favors in both models $d$-wave symmetry, while the extended $s$-wave is stable for more negative values of $\mu$.}
	\label{FS}
\end{figure}

We will concentrate on two sets of parameters,\cite{Radtke:1994,Berthod:2017} as summarised in Table~\ref{table:tB-models}.
The normal state characteristics of the two bandstructures are shown in Fig.~\ref{FS}.
In Fig.~\ref{FS}(a) and Fig.~\ref{FS}(b) we show the bandstructures as density plots, with the Fermi surfaces indicated by black lines.
The Fermi velocities vary around the Fermi surfaces as indicated by the arrows.
Clearly, parameter set \#1 leads to an almost circular Fermi surface, while parameter set \#2 gives a more realistic model of a typical high-temperature superconductor.

The normal state density of states is obtained through
\begin{equation}
	{\cal N}_{\mathrm{N}}(E) = \frac{1}{N_k} \sum_{\vk\in 1.\mathrm{BZ}} \delta(E-\xi_{\vk}),
\end{equation}
where $N_k$ is the number of $\vk$-points we include in the first Brillouin zone (1.BZ).
The resulting bulk density of states for the two models are shown in Fig.~\ref{FS}(c).
We outline in Appendix~\ref{Appendix:FS} how these Fermi surface parameters are fed into the quasiclassical calculations.

%%%
%%%

\subsubsection{Superconducting state}

%%%
\begin{table}
	\caption{Superconducting characteristics of the two tight-binding models. All energies are measured in units of $t$. $T^*$ is extracted from calculations of a diamond shaped sample of the stated number of sites, see Section~\ref{sec:results}.}
	\begin{tabular}{|c|c|c|c|c|c|c|c|}
		\hline
		\hline
		\# & $V_d$ & $T_{\mathrm{c}}$ & $\Delta_{d,0}$ & $\xi$/a & $T^*$ & $T^*/T_{\mathrm{c}}$ & \# sites \\
		\hline
		\hline
		1  & 1.4   & 0.118 & 0.147          & 4.84    & 0.016 & 0.14      & 10255    \\
		\hline
		2  & 1.279 & 0.114 & 0.145          & 5.26    & 0.013 & 0.11      & 10255    \\
		\hline
		\hline
	\end{tabular}
	\label{table:tB-SC}
\end{table}
%%%

%%%
\begin{figure*}
	\includegraphics[width=\textwidth]{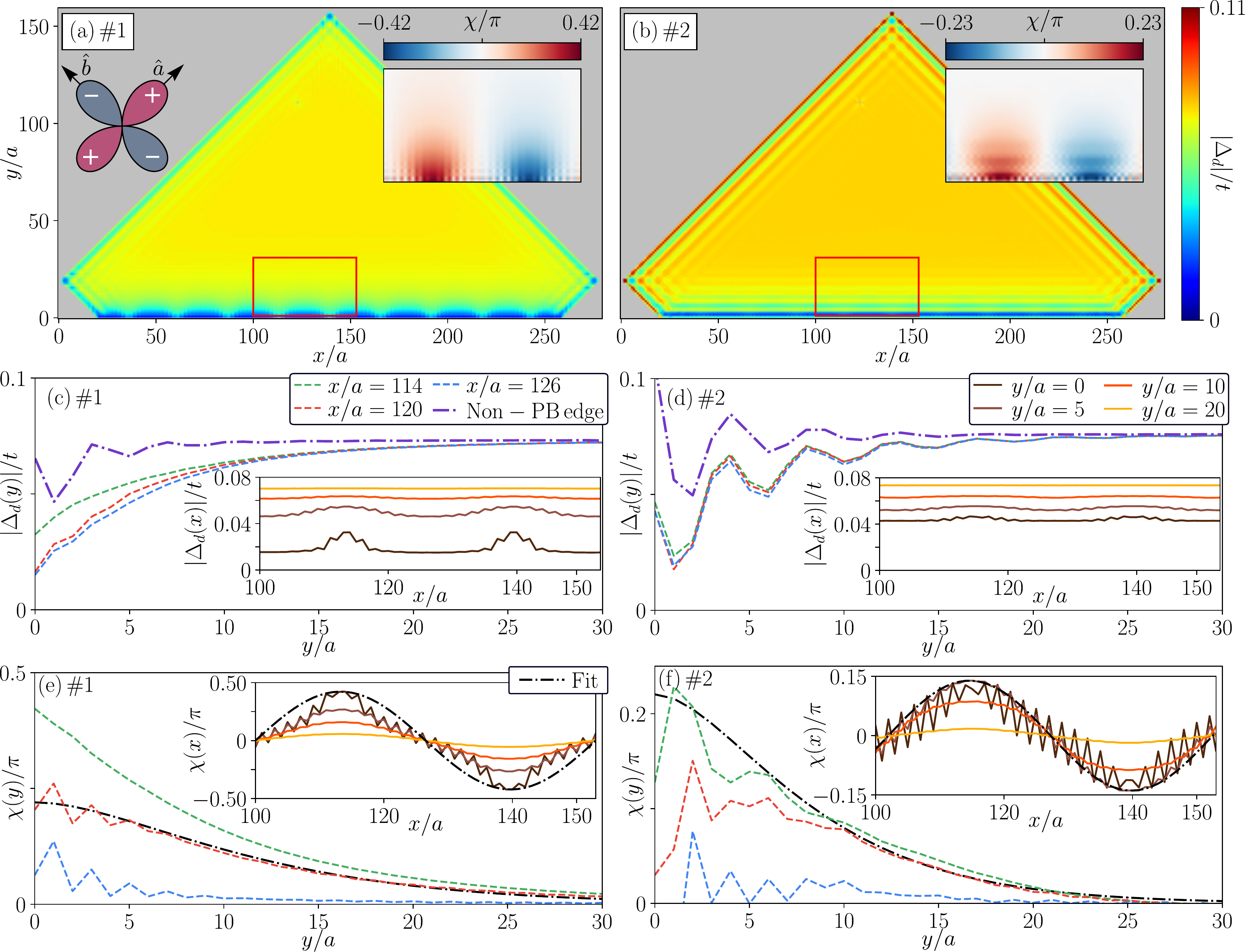}
	\caption{The link order parameter $\Delta_{ij}(\vR)$ for a triangular grain with 12621 sites. The left and right columns contain results for tight-binding models \#1 and \#2, respectively.
		The amplitude $|\Delta_{ij}(\vR)|$ across the whole grain is shown in (a) and (b). The insets contain the phase $\chi_{ij}(\vR)$ within the red squares drawn in the main panels.
		Panels (c)-(d) show the amplitude $|\Delta_{ij}(\vR)|$ as function of $y$ measured away from the grain edges, taken at positions $x$ listed in (c).
		The insets show instead the amplitude as function of $x$ along the edge, taken at positions $y$ listed in (d).
		In (e)-(f) we display $\chi_{ij}(\vR)$ for the same coordinates, where dash-dotted lines are a fit to Eq.~(\ref{eq:phase_crystal_shape}).
		Note that the crystal $\ua$- and $\ub$-axes have been rotated by 45$^{\circ}$ as compared with Fig.~\ref{lattice}.
	}
	\label{fullgrain}
\end{figure*}
%%%

Turning to the  bulk superconducting state, we focus on the $d$-wave, i.e. the  $B_1$ link order parameter.
The pairing interaction and order parameter are
\begin{eqnarray}
	V_{\vk,\vk'} &=& V {\cal Y}_d({\vk}) {\cal Y}_d(\vk'),\\
	\Delta_{\vk}     &=& \Delta_d {\cal Y}_d({\vk}),
\end{eqnarray}
where the $d$-wave orbital basis function is
\begin{equation}
	{\cal Y}_d({\vk}) = \cos(k_xa)-\cos(k_ya).
\end{equation}
The normalization is chosen as
\begin{eqnarray}
	\frac{1}{N_k}\sum_{\vk\in 1.\mathrm{BZ}} {\cal Y}_d^2({\vk}) = 1.
\end{eqnarray}
The temperature-dependent gap equation takes the form
\begin{equation}
	\Delta_d = \frac{V}{N_k} \sum_{\vk\in 1.\mathrm{BZ}} \frac{ \Delta_d{\cal Y}_d^2({\bf k}) } {2E_k} \tanh\left(\frac{E_{\bf k}}{2T}\right),
\end{equation}
where $E_{\bf k} = \sqrt{ \xi_{\vk}^2 + \Delta_{\vk}^2}$. The superconducting coherence length can be defined in different ways, but we will use
\begin{equation}
	\xi/a = \frac{\langle |\vv_{\vkF}|\rangle_{\mathrm{FS}}}{\pi\Delta_{d,0}}
\end{equation}
where $\Delta_{d,0}=\Delta_{d}(T\rightarrow 0)$. The Fermi surface average $\langle ...\rangle_{\mathrm{FS}}$ is defined in Appendix~\ref{Appendix:FS}.
The parameters describing the superconducting state for the different tight-binding realisations are  listed in Table \ref{table:tB-SC}.
For the model parameters we study in this paper, the bulk order parameter symmetry is $d$-wave,
see the variation of $T_{\mathrm{c}}$ with $\mu$ for the two tight-binding models in Fig.~\ref{FS}(d).
The $A_1$ extended $s$-wave channel has a basis function ${\cal Y}_s({\bf k})=\cos(k_xa)+\cos(k_ya)$
and is stable for negative values of $\mu$ in Fig.~\ref{FS}(d). If we would consider this order parameter symmetry as well in the comparison
with a quasiclassical calculation, we should feed in the strength of that subdominant order through its bulk $T_{\mathrm{c}}$.\
We see in Fig.~\ref{FS}(d) that it is zero in model \#1 and $\sim 10^{-3}$ in model \#2 and can therefore be neglected.
Finally, we utilize the relevant Fermi surface basis function ${\cal Y}_d(\vkF)$ in the quasiclassical calculations in Section~\ref{sec:comparison}.

%%%%%%%%%%%%%%%%%
\section{Results: tight-binding model}\label{sec:results}
%%%%%%%%%%%%%%%%%

We have numerically solved the Bogoliubov-de~Gennes equation, Eq.~(\ref{BdGequation}), by direct diagonalization to extract eigenvalues and eigenvectors
for each guess of the link-order parameter in Eq.~(\ref{Delta_link}).
The procedure is iterative and continues until self-consistency between the order parameter and the eigenvalues and eigenvectors has been achieved.
We have studied a range of grain sizes and shapes. For larger systems we have also used the corresponding Green's function formalism
with the same tight-binding Hamiltonians. In this case, we use both the recursive
method and the Chebyshev polynomial expansion method\cite{Covaci:2010}.
For the recursive method, we use our own implementation\cite{Bergvall:2013} of the knitting algorithm\cite{Kazymyrenko:2008}.
All these methods give the same final result.

A subtle detail in the numerics is the initial guess for the order parameter that is needed to find the minimum of the free energy. By guessing a purely real order parameter, the metastable phase without spontaneous currents is always found. We have utilized a few different strategies in order to stabilize the regular pattern of circulating currents. One is to use the ansatz for the phase in Eq.~(\ref{eq:phase_crystal_shape}) below. Another one is to include for the first few iterations an on-site $s$-wave order parameter with a phase winding along the edge and then throw it away. The link order parameter then picks up the phase oscillations. Yet another possiblity is to have a region near the edge where the link order parameter is guessed to have a Larkin-Ovchinnikov\cite{Larkin:1964} type of amplitude oscillation, but with equal magnitudes of the real and imaginary parts. These guesses give different paths to the free energy minimum. In fact, depending on the period of the guessed oscillations, one may stabilize different numbers of current loops in the grain. But there is only one solution that has minimal free energy. We note that a random seed of a complex part of the order parameter results in a lot of noise in the system and it is much harder to find the correct free energy minimum. To summarize, it is important to carefully check for good convergence of the order parameter such that the free energy is minimized and the current is conserved. For the results we show here, current conservation as in Eq.~(\ref{eq:current_conservation}) is upheld to a relative accuracy $\sim 10^{-8}$ at all individual sites.

\begin{figure}[t]
    \includegraphics[width=\columnwidth]{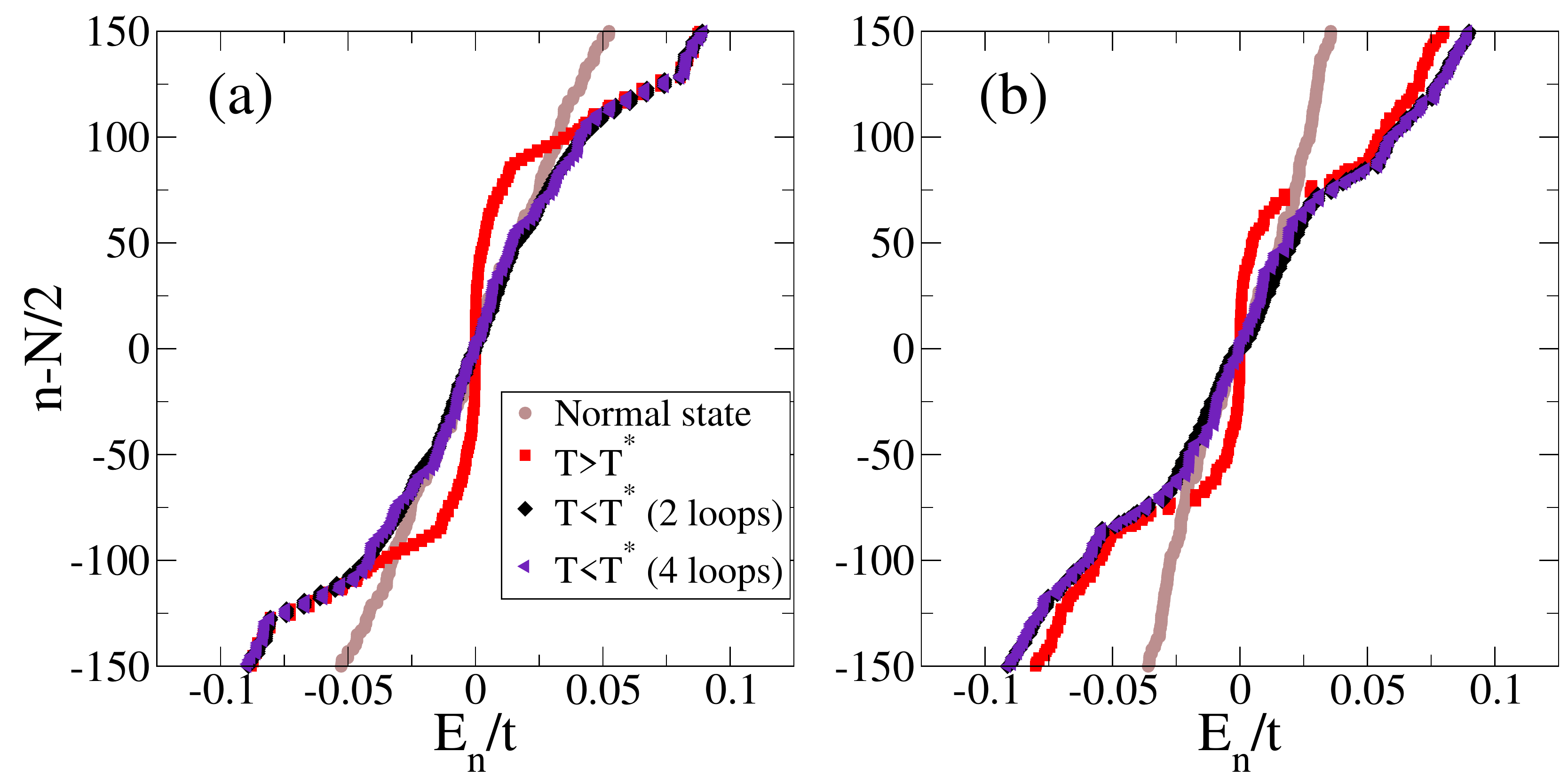}
    \includegraphics[width=\columnwidth]{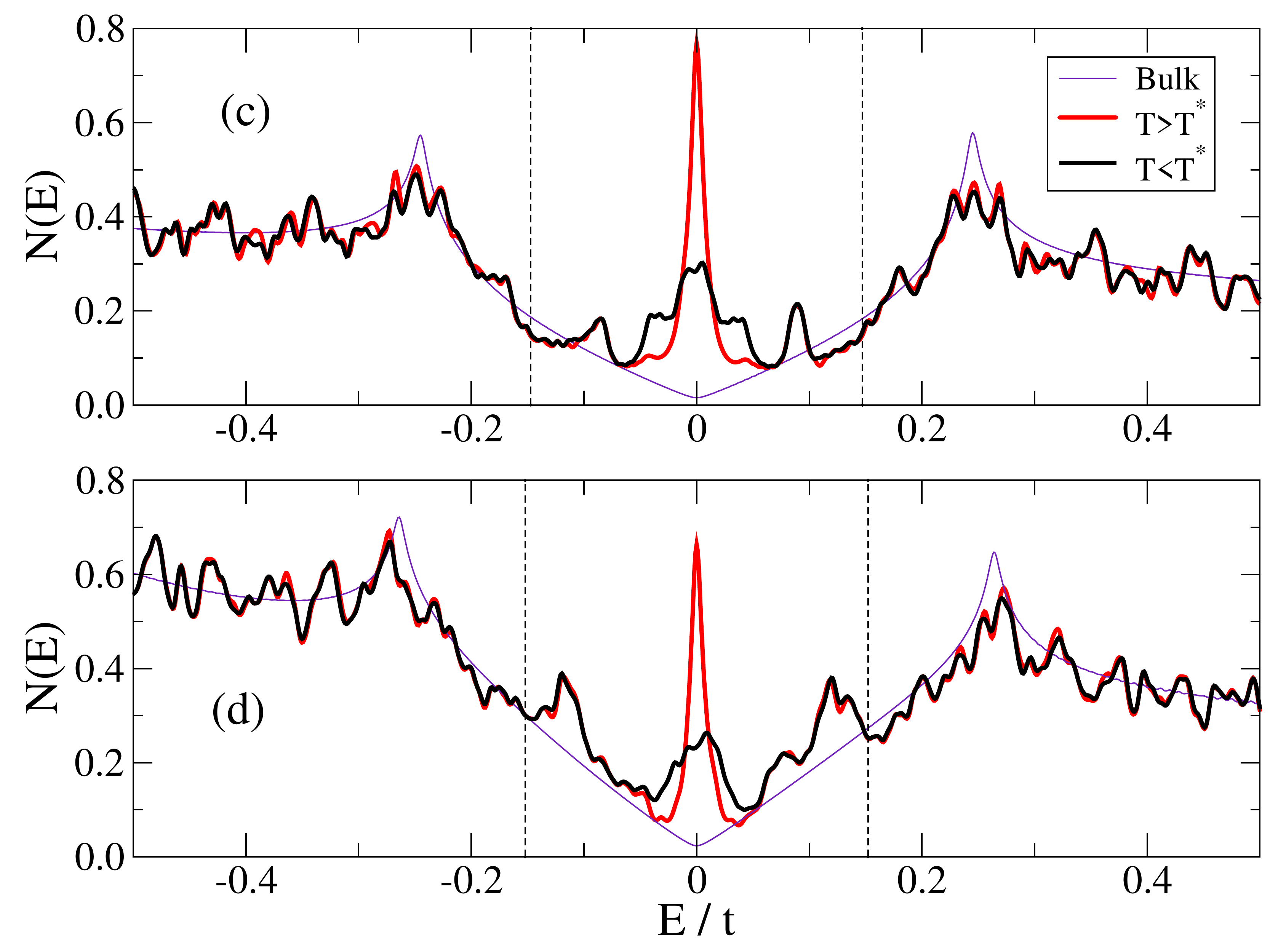}
    \caption{(a)-(b) The distribution of eigenvalues close to the Fermi level $(E=0)$ for the two tight-binding models. The brown circles are eigenvalues for the normal state, while the red squares are for the superconducting state at a temperature above $T^*$. The black diamonds and indigo triangles are for the low-temperature phase with spontaneous loop currents, either with two or four loops per [110] edge. For $T<T^*$ the flat band of eigenvalues are shifted away from the Fermi-level which results in a significant decrease of the free energy in Eq.~(\ref{free-energy}). (c)-(d) The total density states in the superconducting state for the two tight-binding models. The thin indigo lines are for a bulk superconductor. The two other traces are computed for a grain at different temperatures; red line is above $T^*$ and black line is below $T^*$.}
    \label{fig:eigenvalues_dos}
\end{figure}

We show the link order parameter $\Delta_{ij}(\vR)$ for a triangular grain with 12621 sites in Fig.~\ref{fullgrain}.
Results for tight-binding models \#1 and \#2 are displayed in the left and right columns, respectively.
The grain has a single pair breaking [110] edge at $y=0$.
At the [110] edge the well-known zero-energy states are formed\cite{Hu:1994,Kashiwaya:2000,Lofwander:2001}.
In the tight-binding model they show up as a large number of eigenvalues at zero energy, see red squares in Fig.~\ref{fig:eigenvalues_dos}(a)-(b).
The corresponding eigenvectors have large amplitudes at the pair breaking edge, and the order parameter is suppressed there.
This flat band of zero-energy states is energetically unfavorable, and as discussed in the introduction a phase transition can be induced at $T^*$
where spontaneous current loops appear at the edge. Such currents create phase gradients and superflow that Doppler shifts the zero-energy
states away from zero energy thereby lowering the free energy, see also Section~\ref{sec:thermodynamics} below.
The shifts in the energy eigenvalues are also clearly seen in Fig.~\ref{fig:eigenvalues_dos}(a)-(b), black diamonds and indigo triangles.
The resulting total density of states of the grain are shown for both models in Fig.~\ref{fig:eigenvalues_dos}(c)-(d).
The large zero energy peak for $T>T^*$ due to the flat band of zero-energy eigenvalues is clearly broadened for $T<T^*$.
At the same time, the high energy spectra, including the oscillations due to the finite size of the grain, remain unchanged when lowering the temperature from above to below $T^*$.

For $T<T^*$ the order parameter phase $\chi_{ij}$ acquires oscillations along the [110] edge with a period $\sim 50a\sim 10\xi$ 
i.e. a few superconducting coherence lengths to fit a pair of counter-flowing loop currents, see Fig.~\ref{fullgrain}(e)-(f).
We introduce the wavevector $1/q_x\sim\xi$ of this oscillation to adapt to the notation in Ref.~\onlinecite{Holmvall:2020}, where a variational ansatz was introduced for the phase oscillations near $T^*$ within quasiclassical theory. It is
\begin{equation}
	\label{eq:phase_crystal_shape}
	\chi(x,y) \propto \left(1+\frac{y}{y_0}\right)e^{-y/y_0}\cos(q_xx),
\end{equation}
and fits to the tight-binding results are plotted as black dash-dotted lines in Fig.~\ref{fullgrain}(e)-(f).
Near the edge, the amplitude of the phase oscillations is rather large, of the order $\pi$, but there is no phase winding and no topological defects in the order parameter.
The phase decays to zero away from the edge on the scale $y_0\sim\xi$.

The amplitude $|\Delta_{ij}(\vR)|$ also acquires a small oscillation parallel to the edge on the same scale $1/q_x$, although this effect is more pronounced in tight-binding model \#1 than in tight-binding model \#2.
The additional fast oscillations on the scale of the lattice constant $a$ are inherent to the tight-binding model.
There are also oscillations of the amplitude on the scale of the inverse Fermi wave vector $1/k_{\mathrm{F}}\sim 3a$ to $4a$ when looking at the amplitude as a function
of distance from the edges. These also appear at the non-pairbreaking edges,\cite{Zhang:2012,Zhang:2013}
see indigo dash-dotted lines in Fig.~\ref{fullgrain}(c)-(d).
These oscillations are a result of Friedel oscillations at the edge and are temperature independent.

%%%
\begin{figure}
	\includegraphics[width=\columnwidth]{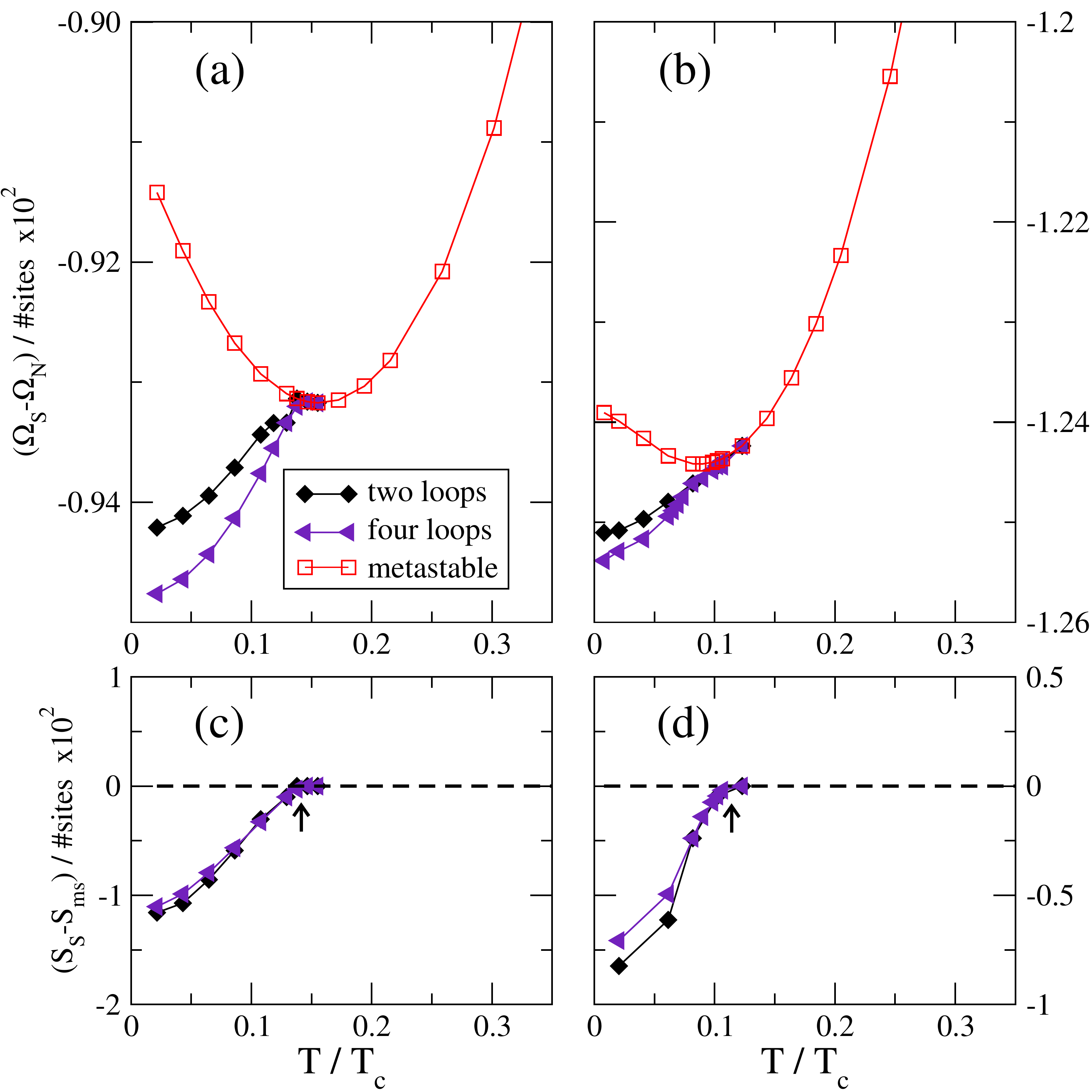}
	\caption{Thermodynamics of a square $d$-wave~superconducting grain with 10255 sites with all four edges pair breaking.
	The left column shows results for tight-binding model \#1, while the right column shows results for tight-binding model \#2.
	In (a)-(b) we show the free energy difference between the superconducting and normal states, $\Omega_{\mathrm{S}}-\Omega_{\mathrm{N}}$,
	normalized to the number of sites.
	The red open squares are results for a purely real order parameter with no spontaneous currents, which is a metastable state for $T<T^*$.
	The black diamonds and indigo triangles are results with spontaneous currents, where the black (indigo) indicates two (four) current loops stabilized along each edge.
	The upturn at low temperatures for the metastable (ms) state is due to the edge Andreev bound states at zero energy, which cost free energy.
	The most stable state is the one with four current loops (indigo triangles) for both models.
	In (c)-(d) we display the difference in entropies between the states with and without spontaneous currents,
	$\mathcal{S}_{\mathrm{S}}-\mathcal{S}_{\mathrm{ms}}$.
	The arrows indicate $T^*$. Note that we have normalized the temperature scale by the two different critical temperatures $T_{\mathrm{c}}$ of the tight-binding models. 
	}
	\label{fig:phasediagrams}
\end{figure}
%%%

\subsection{Thermodynamics}\label{sec:thermodynamics}

We present in Fig.~\ref{fig:phasediagrams} the low-temperature thermodynamics of a square $d$-wave superconducting grain
with all four edges at a 45$^\circ$ angle from the $ab$-axes.
This grain has 10255 sites and is slightly smaller than the one in Fig.~\ref{fullgrain}.
In particular, each edge is much shorter than in the triangular grain.
The free energy of the metastable phase, which has no currents and a purely real order parameter [red open squares in Fig.~\ref{fig:phasediagrams}(a)--(b)], shows a pronounced upturn at low temperature, signalling the energy cost of the zero-energy Andreev bound states. For larger grains, the upturn is less visible since the weight of the edge is diminished compared with the grain interior.
It is interesting that we can stabilize different number of current loops along each edge: i) four current loops, purple triangles, and ii) two current loops, black diamonds.
The two configurations have the same $T^*$. For lower temperature, $T<T^*$, the configuration with four current loops (purple triangles) has lower free energy.
From the entropy, Fig.~\ref{fig:phasediagrams}(c)-(d), one can see from the abrupt change of slope at $T^*$ that the phase transition is of second order
and has the same value within our numerical accuracy irrespective of number of current loops.
To enhance the visibility of the knee in the entropy as function of temperature, we present the difference in entropy between the phase with currents and
the metastable phase which has no currents, ${\mathcal S}_{\mathrm{S}}-{\cal S}_{\mathrm{ms}}$.

The difference in $T^*$ between the two tight-binding models can be understood as due to the different number of zero-energy eigenvalues. In Fig.~\ref{fig:eigenvalues_dos} we see that there are more eigenvalues and a higher spectral weight in tight-binding model \#1. The Doppler shifts then become more energetically favorable, and consequently, $T^*$ is higher, see Table~\ref{table:tB-SC}. The higher energy cost of the zero-energy states in model \#1 is also visible in the larger upturn in the free energy of the metastable state, see Fig.~\ref{fig:phasediagrams}(a)-(b).

Thus, we can conclude that the phase crystallization that was studied in Refs.~\onlinecite{Hakansson:2015,Holmvall:2018a,Holmvall:2018b,Holmvall:2019,Holmvall:2020}
within the quasiclassical approximation can be found also in the more general mean field tight-binding Bogliubov-de~Gennes theory.
Early work\cite{Zhu:1999,Nagai:2017} on this type of tight-binding model also found structures in the order parameter and spontaneous currents.
At that time only small grains could be studied, probably because of limited computational resources, and the regular amplitude and phase oscillations that we find here were not seen.
In those works, the appearance of edge currents was associated with the development of the $A_1$ site quantity in Eq.~(\ref{site_Delta_s}).
Here, instead, we associate the spontaneous currents with phase crystallization and identify the resulting Doppler shifts
as the microscopic mechanism behind the lowering of the free energy below the phase transition temperature $T^*$.
We also note that, at least from quasiclassical theory, a subdominant $s$-wave order parameter at the edge would favor a translational invariant current flow along each edge, see Ref.~\onlinecite{Hakansson:2015}. In addition, for a subdominant $s$-wave order parameter we would expect a spectral gap around zero energy, which is not present in Fig.~\ref{fig:eigenvalues_dos}.
To further support these conclusions, we compare in the next section with results that we have obtained with the quasiclassical theory.

\section{Results: comparison with quasiclassical theory}
\label{sec:comparison}

% Add Anton reference
The quasiclassical theory of superconductivity\cite{Serene:1983,eschrig:2000} is used to study triangular grains, following the methods described in Refs.~\onlinecite{Hakansson:2015,Holmvall:2018a,Holmvall:2018b,Holmvall:2019,Holmvall:2020}, but with modified Fermi-surface averages\cite{Buchholtz:1995} as explained in App.~\ref{Appendix:FS}. It is noted that Fermi-surface effects were studied\cite{Miyawaki:2015,Miyawaki:2017} for a similar spontaneous-TRSB phase\cite{Vorontsov:2009,Hachiya:2013}. In addition to the Fermi surfaces defined by tight-binding models \#1 and \#2, a circular Fermi surface is also investigated. It shows very similar results to model \#1 due to the similar Fermi surface shape, and is therefore omitted from some of the figures. We begin by comparing the self-consistent order-parameter and current obtained with quasiclassics versus with BdG. This is followed by an analysis of the thermodynamics and the phase transition, where numerical and analytical calculations are combined to explain the results in terms of the Fermi-surface features.

\subsection{Self-consistent observables}\label{sec:qc:observables}
\begin{figure*}
	\includegraphics[width=\textwidth]{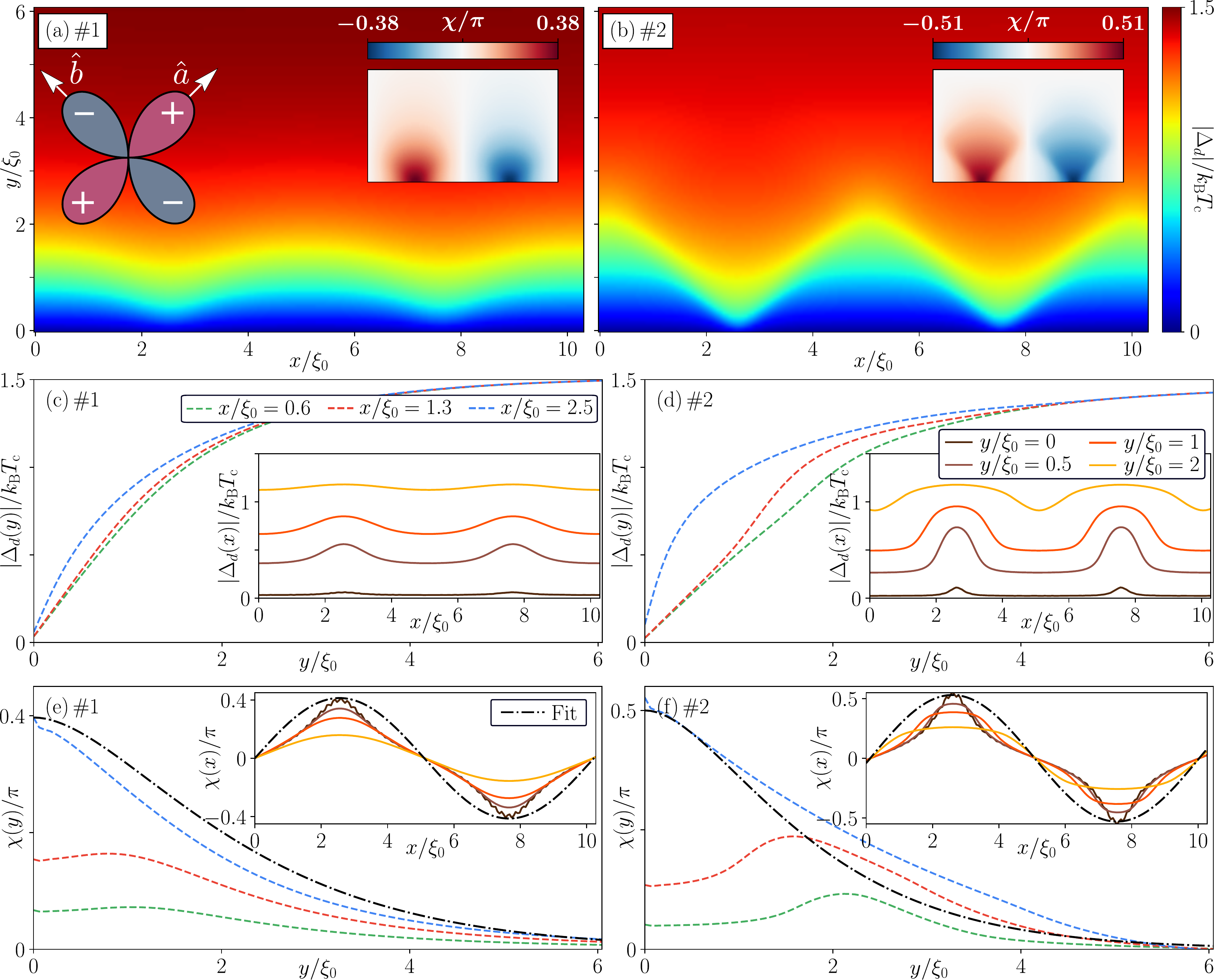}
	\caption{The quasiclassical order parameter $\Delta_d(\mathbf{R}) = \left|\Delta_d(\mathbf{R})\right|\exp\left({i\chi(\mathbf{R})}\right)$ at $T=0.1T_{\mathrm{c}}$ in a triangular grain with a single pair breaking edge of length $60\xi_0$ at $y=0$. The left and right columns contain results for tight-binding models $\#1$ and $\#2$, respectively. Panels (a) and (b) show the amplitude $\left|\Delta_d(\mathbf{R})\right|$ in a unit-cell at the center of the pair-breaking edge, with insets showing the phase $\chi(\mathbf{R})$. Panels (c)--(d) show the amplitude as a function of $y$ measured perpendicular to the interface, taken at positions $x$ listed in (c). The insets show instead the amplitude as function of $x$ measured parallel to the edge, taken at positions $y$ listed in (d). Panels (e)--(f) show the phase $\chi(\mathbf{R})$ for the same coordinates, with a fit to Eq.~(\ref{eq:phase_crystal_shape}). The fit is perfect at $T \lesssim T^{*}$, but at lower temperatures, the figures show that there is a clear deviation from the fit due to higher-order modulations. }
	\label{fig:qc_order_parameter}
\end{figure*}

\begin{figure*}
	\includegraphics[width=\textwidth]{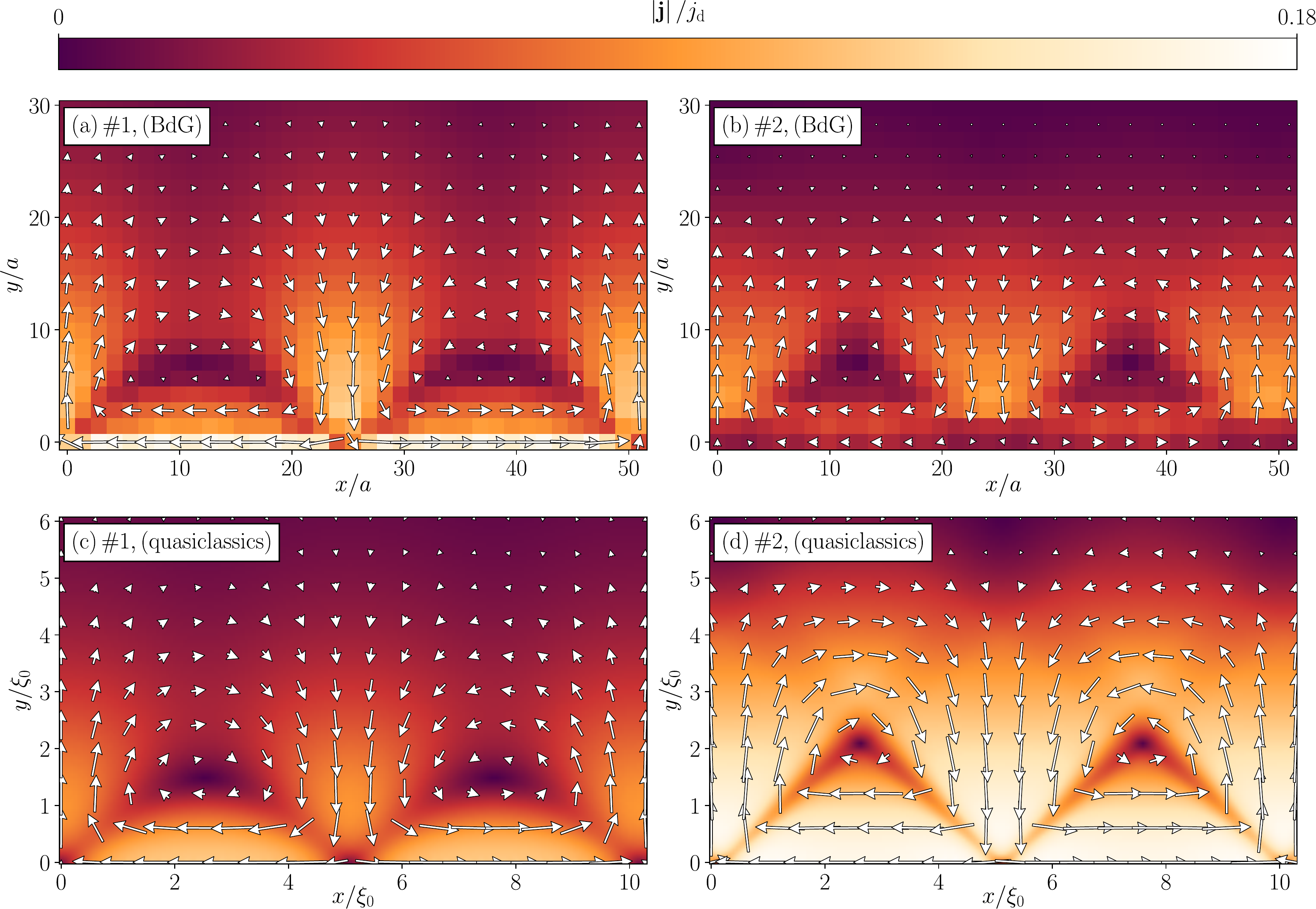}
	\caption{Circulating currents near the [110] edge, using (a)--(b) BdG at temperature 
	$T = 0.005t\approx 0.04T_{\mathrm{c}} < T^*$, and (c)--(d) quasiclassics at temperature $T = 0.1T_{\mathrm{c}} < T^*$. The left and right columns are for tight-binding models \#1 and \#2, respectively. Here, the depairing current is defined as $j_{\mathrm{d}} \equiv 4\pi T_{\mathrm{c}}|e|\NF\barvF$.
	}
	\label{fig:currents}
\end{figure*}

As the system enters the symmetry-breaking phase, all three models follow Eq.~(\ref{eq:phase_crystal_shape}), with a phase that varies sinusoidally with wavenumber $q_x \sim 1/\xi_0$ parallel to the interface, and that decays exponentially over distance $y_0\sim\xi_0$ away from the interface. Here, we use the definition $\xi_0 \equiv \langle|\vvF|\rangle\FS/(2\pi T_{\mathrm{c}})$. Such a superflow gives rise to smooth and round currents \cite{Hakansson:2015,Holmvall:2018b,Holmvall:2020}, very similar to those found with the BdG approach. As the temperature is lowered far below the transition temperature, however, the shape of the phase is significantly modified, from sinusoidal to triangle-wave-like. This ensures long and constant phase gradients, and hence constant superflow $\mathbf{p}_{\mathrm{s}}$, which leads to the most efficient Doppler shift of midgap states. This is seen in the deviation from the fit to Eq.~(\ref{eq:phase_crystal_shape}) in Fig.~\ref{fig:qc_order_parameter}, where the order-parameter amplitude and phase are plotted at $T=0.1T_{\mathrm{c}} < T^*$, with results for models $\#1$ and $\#2$ displayed in the left and right columns, respectively. The same deviation from Eq.~(\ref{eq:phase_crystal_shape}) seems to be somewhat present in the BdG results as well, but is highly obscured by the atomic-scale oscillations parallel to the interface (see Fig.~\ref{fullgrain}).
The resulting currents are qualitatively very similar in quasiclassics and BdG, see Fig.~\ref{fig:currents}.

It is noted that the size of the unit cell illustrated in the BdG results is roughly $50a \times 30a$, which with a coherence length of $\xi \sim 5a$ corresponds to the unit cell of roughly $10\xi_0 \times 6\xi_0$ illustrated in the quasiclassical results. Furthermore, the same peculiar pattern of edge-defects (i.e. sources, sinks and saddle-points) in the superflow vector field $\mathbf{p}_{\mathrm{s}}$ as reported in Ref.~\onlinecite{Holmvall:2018b} are still present in all Fermi surface models (not shown here).

%%%
%%%

\subsection{Thermodynamics of the phase transition}\label{sec:qc:thermodynamics}
\begin{figure}[t]
	\includegraphics[width=\columnwidth]{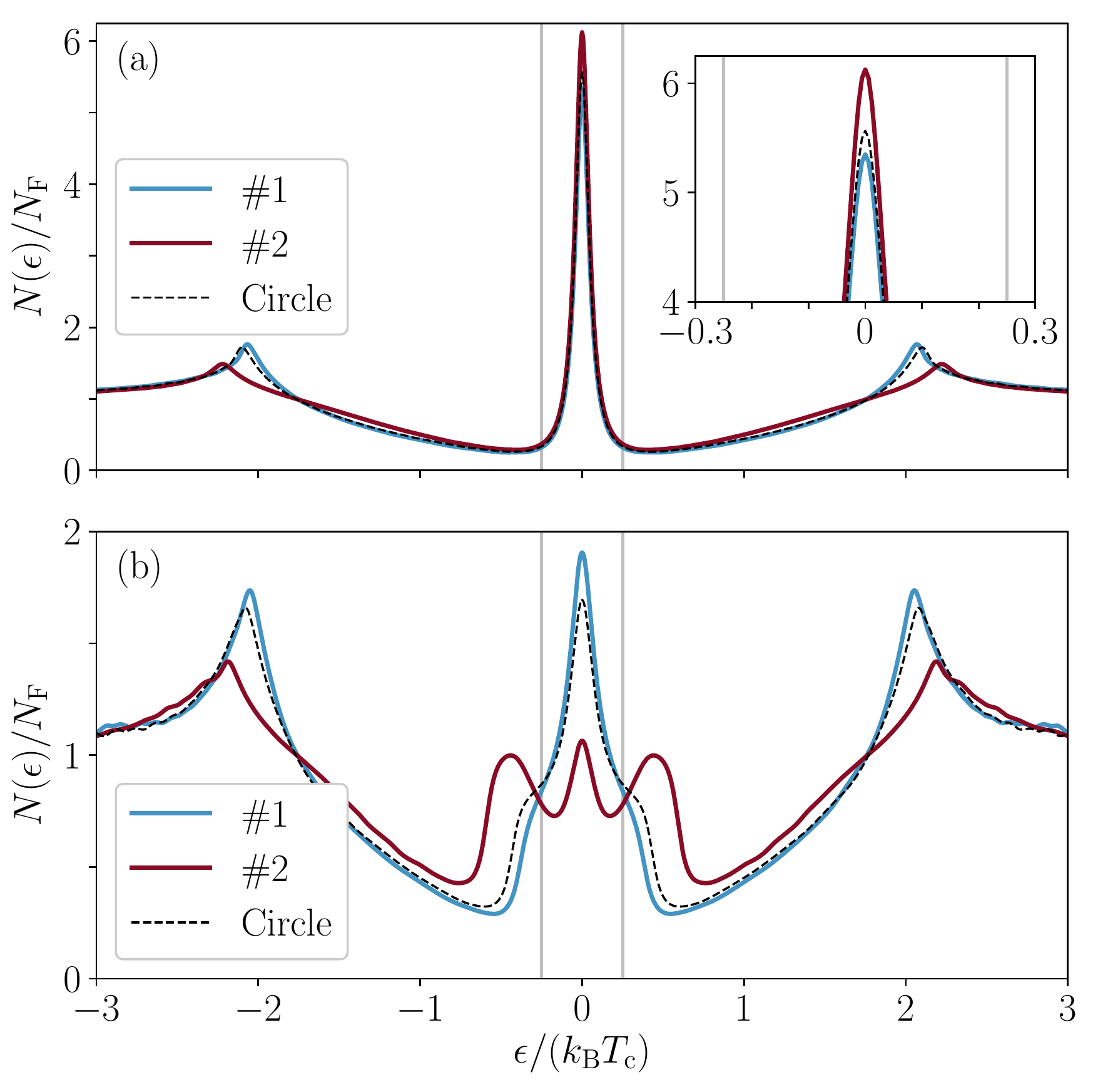}
	\caption{The average density of states, $N({\bf R}; \epsilon)$, at the pair-breaking interface and temperature $T=0.1 T_\mathrm{c} < T^*$, (a) in the meta-stable phase without spontaneous currents, and (b) in the symmetry-broken phase. The average is taken within one superflow period ($\sim 10\xi_0 \times  6\xi_0$) at the interface, as shown in Fig.~\ref{fig:qc_order_parameter} (a)--(b). The gray lines mark the integration interval of the spectral weight, Eq.~(\ref{eq:spectral_weight}). The inset of (a) shows the zero-energy peak within this interval.}
	\label{fig:qc_ldos}
\end{figure}

The pair-breaking interface leads to zero-energy midgap states (MGS) with an associated energy cost, in the form lost condensation energy, that scales as $1/T$. In the presence of a superflow $\mathbf{p}_{\mathrm{s}}$, the zero-energy states are Doppler shifted to finite energies $\delta_{\epsilon}\propto\vvF\cdot\mathbf{p}_{\mathrm{s}}$, consequently reducing the energy cost. However, the superflow also shifts the continuum states in the bulk, where it instead costs kinetic energy. The optimal form of the superflow is therefore an exponential decay away from the pair-breaking interface\footnote{Superflow with functional forms that decay faster into the bulk than exponential decay lead to a higher free energy, due to gradient terms connected with the off-diagonal terms in the superfluid-density tensor\cite{Holmvall:2020,Holmvall:2019b}.}, as in Eq.~(\ref{eq:phase_crystal_shape}). Due to the $1/T$-dependence of the MGS energy cost, the energy gain of spontaneous superflow at the interface eventually outweighs the cost of its tail in the bulk, below a transition temperature $T^*$. This transition temperature is extracted from the self-consistent numerics via the quasiclassical free energy\cite{Virtanen:2020}, by comparing the free energy of systems with and without spontaneous superflow. The latter is referred to as meta-stable (ms) as it exhibits a higher free energy below $T^*$.
\begin{table}[t!]
	\caption{Quasiclassical results obtained with full self-consistent numerics ($T^*$ and $A$), and obtained by evaluating analytic low-temperature expressions ($\Delta$ and $\delta\Omega$, see App.~\ref{sec:qc_lowtemp_analytic}), for the three different Fermi surface models. Here, $A$ and $A_{\mathrm{ms}}$ are the spectral-weights with superflow and in a meta-stable (ms) system without superflow, respectively, integrated over one superflow period ($\sim 10\xi_0 \times  6\xi_0$) at the interface following Eq.~(\ref{eq:spectral_weight}), and evaluated at temperature $T=0.1T_\mathrm{c} < T^*$. The spectral energy range is $\epsilon/T_\mathrm{c} \in [-5\eta,5\eta]$, where $\eta = 0.05$ is the smearing (i.e. magnitude of the small imaginary part added to the energy when computing the LDOS), and $A_0 = T_\mathrm{c} N_\mathrm{F}$. For the analytic results, $\Delta_0^{\mathrm{bulk}}$ denotes the bulk gap, $\delta\Omega^{\mathrm{BCS}}$ the BCS free energy (i.e. without any pair-breaking interface), $\delta\Omega^{\mathrm{MGS}}$ the energy-cost of the midgap states, and $\delta\Omega^{\mathrm{DS}}$ the energy-gain of Doppler shifting the midgap states to finite energies. Finally, the energies and superflow are given in units $\Omega_{\mathcal{A}} \equiv \mathcal{A}\NF  T_{\mathrm{c}}^2$, $\Omega_{\mathcal{L}} \equiv \mathcal{L}\xi_0\NF T_{\mathrm{c}}^2$ and $p_0 \equiv T_{\mathrm{c}}/\barvF$, with $\mathcal{A} \equiv \int dxdy$, $\mathcal{L} = \int dx$ and $\barvF \equiv \langle\left|\mathbf{v}_{\mathrm{F}}\right|\rangle\FS$.}
	\begin{tabular}{| c | c | c | c |}
		\hline
		\hline
		Fermi surface                                                             & \#1                                             & \#2                                             & Circular                                       \\
		\hline
		\hline
		$T^*/T_{\mathrm{c}}$                                                      & $0.15$                                          & $0.25$                                          & $0.18$                                         \\
		\hline
		$A_\mathrm{ms} / A_0$                                                     & $0.75$                                          & $0.85$                                          & $0.78$                                         \\
		\hline
		$A / A_0$                                                                 & $0.60$                                          & $0.39$                                          & $0.56$                                         \\
		\hline
		$\Delta_0^{\mathrm{bulk}}/T_{\mathrm{c}}$                   & $1.53$                                          & $1.46$                                          & $1.51$                                         \\
		\hline
		$\delta\Omega^{\mathrm{BCS}}/\Omega_{\mathcal{A}}$                              & $-1.64$                                         & $-1.07$                                         & $-1.14$                                        \\
		\hline
		$\delta\Omega^{\mathrm{MGS}}/\Omega_{\mathcal{L}}$                              & $4.31$                                          & $4.89$                                          & $4.48$                                         \\
		\hline
		$\delta\Omega^{\mathrm{DS}}(\mathbf{p}_{\mathrm{s}})/\Omega_{\mathcal{L}}$ & $-2.76\left|\mathbf{p}_{\mathrm{s}}\right|/p_0$ & $-4.48\left|\mathbf{p}_{\mathrm{s}}\right|/p_0$ & $-\pi\left|\mathbf{p}_{\mathrm{s}}\right|/p_0$ \\
		\hline
		\hline
	\end{tabular}
	\label{table:qc:thermodynamics}
\end{table}
Table~\ref{table:qc:thermodynamics} presents the numerically obtained $T^*$ and average spectral weight ($A$) for all three Fermi-surface models, together with evaluated low-temperature analytic expressions of the gap and free-energy terms (explained further in App.~\ref{sec:qc_lowtemp_analytic}). Model $\#1$ ($\#2$) shows a lower (higher) $T^*$ than that of a circular Fermi surface, which correlates with a lower (higher) cost of the MGS, and a lower (higher) energy gain of Doppler-shifting them. The density of states also follows this trend, with a lower (higher) spectral weight for model $\#1$ ($\#2$) above $T^*$. This is because there are less (more) zero-energy states at the interface, that extend over a shorter (longer) distance into the bulk due to a smaller (larger) suppression of the order-parameter amplitude $|\Delta|$, see Fig.~\ref{fig:qc_ldos} (a). The spectral weight is obtained by integrating the local density of states, $N(\vR; \epsilon)$, in a small window around zero energy, and spatially averaging over one unit cell (u.c.) of the periodic phase (i.e. the region shown in Fig.~\ref{fig:qc_order_parameter}), according to
\begin{equation}
	\label{eq:spectral_weight}
	A = \frac{1}{\mathcal{A}_{\mathrm{u.c.}} } \int_{\mathrm{u.c.}} \mathop{d\vR} \int^{\epsilon_\mathrm{c}}_{-\epsilon_\mathrm{c}} N(\vR; \epsilon)\mathop{d\epsilon} \; ,
\end{equation}
where $\mathcal{A}_{\mathrm{u.c.}} = \int_{\mathrm{u.c.}} \mathop{d\vR}$. Figure~\ref{fig:qc_ldos} shows that in the presence of superflow below $T^*$, the Doppler shift is lower (higher) for tight-binding model $\#1$ ($\#2$) compared to the circular Fermi surface, leading to a higher (lower) spectral weight close to zero energy.
To summarize, model $\#2$ has the highest spectral weight of zero-energy states, and the most efficient Doppler shift, yielding the highest $T^*$. Note that this is in contrast to the BdG results, where model $\#1$ has the highest spectral weight and $T^*$.

The quasiclassical results are further understood by analyzing the dependence on the Fermi velocity, $\vF(\vk_\mathrm{F})$, in the angular averages entering the observables (explained in App.~\ref{Appendix:FS}). The Fermi velocity is constant across the circular Fermi surface, while it varies for models $\#1$ and $\#2$, with maxima and minima which either align with the nodes or the antinodes (lobes) of the $d_{x^2 -y^2}$-wave basis function, as is seen in Fig.~\ref{FS}(a)--(b). Model \#1 has its Fermi velocity maxima along the nodes, and its minima along the antinodes, while it is the opposite\footnote{In polar coordinates with the origin at $\vk = (\pi, \pi)$ the Fermi velocity minimas for model \#2 are situated at $\pm 0.044\pi$ radians off of the diagonals.} for model \#2. The full analysis is non-trivial, but in short, this behaviour ensures that model \#2 and \#1 weighs the MGS contribution higher and lower, respectively (see App.~\ref{sec:qc_lowtemp_analytic} for further details).
This trend was verified by also investigating the cost of the midgap states for a $d_{xy}$-wave basis function with a pair-breaking $[100]$ interface (not shown here). Here, the Fermi-surface shows the opposite features, such that the opposite trend is expected for $T^*$. Indeed, in this case, model \#1 and \#2 have instead a larger and smaller transition temperature than the circular Fermi surface, respectively.

To conclude, the quasiclassical results thus highlight that the zero-energy states and the transition temperature can change quite significantly with the Fermi surface parameters, and that it is possible to pinpoint the dependence on these parameters for various observables and quantities. In the tight-binding results, there are additional microscopic effects complicating the situation, but the end result is the same in the sense that the transition is governed by the spectral weight of zero-energy states.

\section{Summary}
\label{sec:summary}

In summary, we have studied two tight-binding models of $d$-wave superconducting grains with pair-breaking [110] edges. At a phase-transition temperature $T^*$, the order parameter develops phase gradients, and circulating currents appear near the edges. This extends our earlier results\cite{Hakansson:2015,Holmvall:2018a,Holmvall:2018b,Holmvall:2019,Holmvall:2020} within the quasiclassical approximation to a more general theory with a realistic Fermi surface and where fast oscillations on the scale of the Fermi wavelength are taken into account. We find that the phase transition and the qualitative characteristics of the symmetry-broken phase are universal, independent of including the microscopic details or not. On the other hand, some quantitative details, such as the predicted $T^*$, depend on the model. For instance, $T^*$ of tight-binding model \#2 is lower than \#1 within BdG, while it is higher within the quasiclassical calculation. This deviation can be due to several things: e.g. electron-hole asymmetry not included in quasiclassics, or details of the eigenfunctions and Friedel oscillations included in BdG. But as we have shown we can still understand the variation of $T^*$ in all cases (two models within both BdG and quasiclassics) in terms of the spectral weight of zero-energy Andreev bound states. Higher spectral weight leads to a higher $T^*$, since the gain in free energy due to the Doppler shifts then increases. Within quasiclassics, the variation of the Fermi velocity around the Fermi surface is also important. The universality can be understood since the scale of variations of the current patterns and the phase variations is a few coherence lengths, which is much larger than the fast $1/k_{\mathrm{F}}$ oscillations. When fast oscillations, and the high energy parts, are integrated out when making the quasiclassical approximation, the physics behind the second-order phase transition at $T^*$ is kept. Since the tight-binding model is more general than quasiclassical theory, we may assume that the predictions would be more reliable. On the other hand, the tight-binding model we have studied also has its approximations, including the weak-coupling mean field approximation. As an outlook, it would therefore be of interest to go beyond mean-field and include effects of electron correlations.\cite{Matsubara:2020a,Matsubara:2020b}

\section{Acknowledgements}
We thank A. B. Vorontsov for valuable discussions. The numerical computations were performed on resources at Chalmers Centre for Computational Science and Engineering provided by the Swedish National Infrastructure for Computing. We thank the Swedish Research Council (VR) for financial support.

\appendix

\section{Fermi surface average}
\label{Appendix:FS}

Within the quasiclassical theory of superconductivity,\cite{Serene:1983,eschrig:2000}
the calculation of an observable or a self-energy involves a Fermi surface average.
In our case of a two-dimensional system the average is over the Fermi lines in Fig.~\ref{FS}.
Each of them defines a contour $C$.
The average of an arbitrary function $f(\vkF)$ then takes the form of a line integral
\begin{equation}
	\label{eq:fs_average_integral}
	\langle f(\vkF) \rangle\FS = \frac{1}{\NF} \oint_{C} \frac{ds}{(2\pi)^2|\vvF(s)|} f(\vkF(s)).
\end{equation}
The total density of states at the Fermi energy is
\begin{equation}
	\NF =  \oint_{C} \frac{ds}{(2\pi)^2|\vvF(s)|}.
\end{equation}
Let us extend the bandstructures in Fig.~\ref{FS} to a repeated zone scheme, choose origo at $\vk=\left(\pi,\pi\right)$, and use polar coordinates.
For the parameters of our tight-binding models, the contour then becomes a closed non-circular loop that can simply be parameterized by the azimuth
$\varphi\in(-\pi,\pi]$.
The elementary arc length is
\begin{equation}
	ds = \sqrt{\left(\frac{dk_{\mathrm{F}x}}{d\varphi}\right)^2 + \left(\frac{dk_{\mathrm{F}y}}{d\varphi}\right)^2}d\varphi = |\vkF^\prime(\varphi)|d\varphi,
\end{equation}
and the line integral takes the form
\begin{equation}
    \label{eq:fs_average_integral_explicit}
	\langle f(\vkF) \rangle\FS = \frac{1}{\NF} \int_{-\pi}^{\pi} \frac{d\varphi}{2\pi} \frac{|\vkF^\prime(\varphi)|}{2\pi |\vvF(\varphi)|} f(\vkF(\varphi)).
\end{equation}
Thus, we extract $\vkF(\varphi)$ numerically from the bandstructure of the chosen tight-binding model. The Fermi velocity $\vvF(\varphi)$ and the derivative $\vkF^\prime(\varphi)$ can be computed from analytic formulas with $\vkF$ as input.\cite{Buchholtz:1995} We may then compute the total density of states at the Fermi energy $\NF$. Note that when viewing the Fermi surface in the first Brillouin zone, as in Fig.~\ref{FS}, we must translate the extracted Fermi momenta $\vkF$ back from higher Brillouin zones.

To compute observables and self-energies, we need to compute the quasiclassical Green's function $\hat g$,
see the Methods section in Ref.~\onlinecite{Holmvall:2019}. That involves solving first-order differential equations along
trajectories defined by the extracted Fermi velocities $\vvF(\varphi)$, see Eqs.~(13)-(14) in Ref.~\onlinecite{Holmvall:2019}.
It is then important to note that the Fermi momentum $\vkF(\varphi)$ and the Fermi velocity $\vvF(\varphi)$ are not parallel.
At the starting points and end points of the trajectories, i.e. at the edges, trajectories couple through a boundary condition.
We use a specular boundary condition, and limit ourselves to grains with high-symmetry edges, either along crystal axes or 45$^{\circ}$ rotated,
like the $[110]$-edge in Fig.~\ref{lattice}.
This means that a trajectory parameterized by $\varphi$ couples to the same trajectories $\varphi'$ as for a circular Fermi surface.

\section{Analytic quasiclassical expressions at low temperatures}\label{sec:qc_lowtemp_analytic}

\begin{figure}[b]
	\includegraphics[width=\columnwidth]{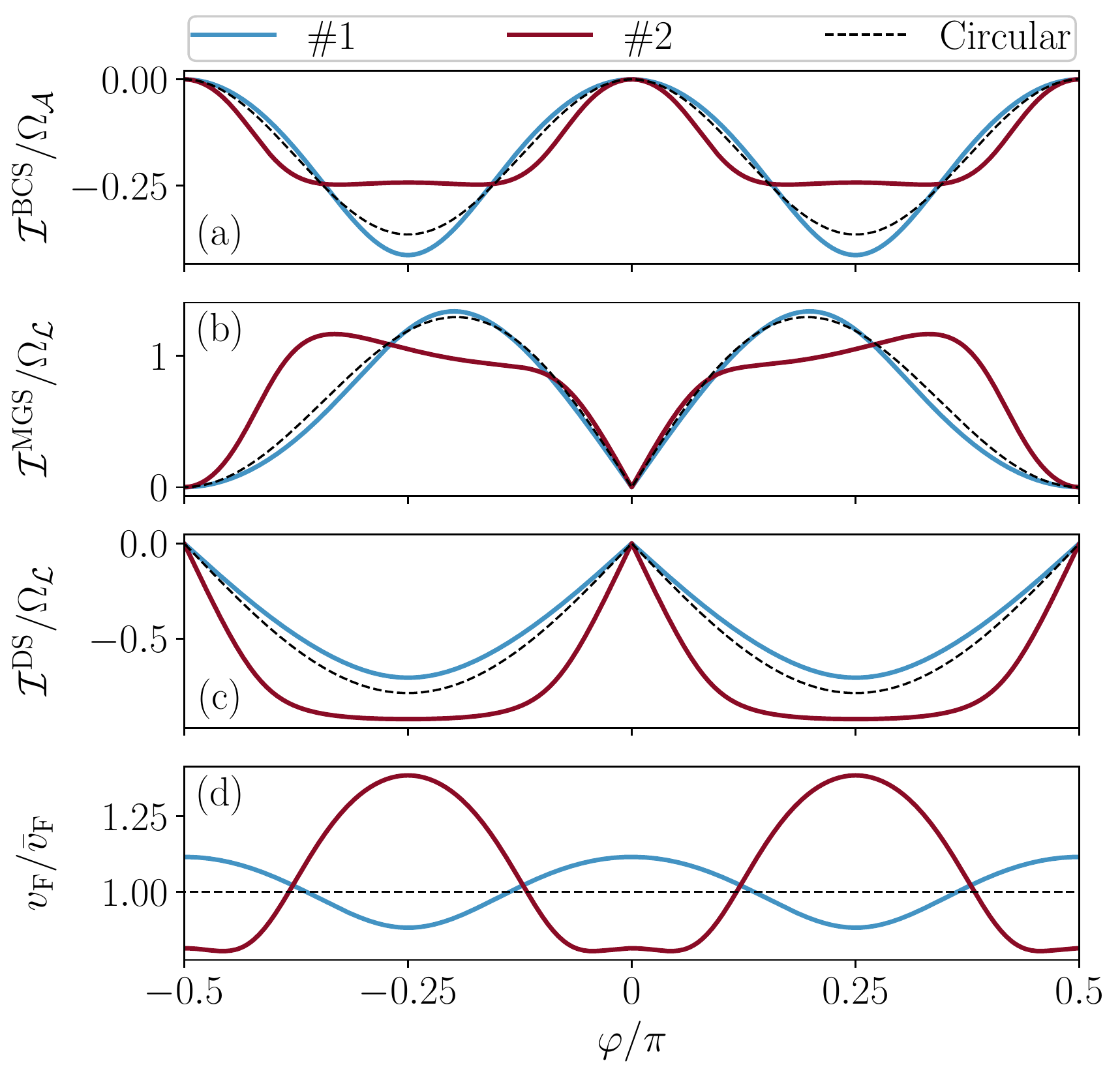}
	\caption{The dependence on the Fermi-surface angle $\varphi$ of the integrands of (a) the BCS free energy from Eq.~\eqref{eq:qc_lowtemp:bcs_energy}, (b) the midgap-state energy cost from Eq.~\eqref{eq:qc_lowtemp:mgs_energy}, (c) the energy gain of Doppler-shifting the midgap states from Eq.~\eqref{eq:qc_lowtemp:doppler_energy}, as well as (d) the Fermi velocity, $\vF(\varphi)$ with $\barvF \equiv \langle\left|\mathbf{v}_{\mathrm{F}}\right|\rangle\FS$. The integrands are defined such that $\Omega^\mathrm{X} = \int^{2\pi}_0\mathcal{I}^\mathrm{X}\, d\varphi $, and the quantities $\Omega_\mathcal{A}$ and $\Omega_\mathcal{L}$ are the same as in Table~\ref{table:qc:thermodynamics}. In this choice of coordinate system the basis-function antinodes (lobes) are situated along $\varphi=(n + 1/2) \pi /2$, and the nodes along $\varphi=n \pi /2$, with $n\in\mathbb{Z}$. The interface normal is parallel to $\varphi=0$.
	}
	\label{fig:qc_analytic_angle_resolved}
\end{figure}

To get an analytic handle on the energetics of the midgap Andreev states, analytic expressions are presented for a bulk $d$-wave superconductor at low temperatures (derived in Ref.~\onlinecite{Holmvall:2019b}). These expressions are used to obtain the gap and the energies in Tab.~\ref{table:qc:thermodynamics}. We use a gague transformation that eliminates the phase of the order parameter in favor of explicit dependence on superflow field $\mathbf{p}_{\mathrm{s}}$ \cite{Holmvall:2018b,Holmvall:2020}, hence rendering the order parameter real. In the absence of superflow, the bulk order parameter at zero temperature can be expressed as $\Delta(\vkF) = \Delta_0 {\cal Y}(\vkF)$ with
\begin{equation}
	\label{eq:qc_lowtemp:order_parameter}
	\Delta_0 = \pi e^{-\gamma_{\mathrm{E}}}e^{-\left\langle|{\cal Y}|^2\ln|{\cal Y}|\right\rangle\FS} T_{\mathrm{c}},
\end{equation}
where $\gamma_{\mathrm{E}}$ is the Euler-Mascheroni constant. For a $d_{x^2-y^2}$ order parameter with circular Fermi surface, the FS integral is taken analytically and reduces to the familiar result $\Delta_0/T_{\mathrm{c}} = \sqrt{2}\pi \exp(-\gamma_{\mathrm{E}}-1/2) \approx 1.51$. For general tight-binding parameters, the FS average takes the form of Eq.~(\ref{eq:fs_average_integral}), and is evaluated numerically for e.g. tight-binding models $\# 1$ and $\# 2$, again, see Tab.~\ref{table:qc:thermodynamics}. Having calculated the gap, the BCS free energy is
\begin{equation}
	\label{eq:qc_lowtemp:bcs_energy}
	\frac{\delta\Omega^{\mathrm{BCS}}}{\mathcal{A}\NF} = -\frac{\langle|\Delta(\vkF)|^2\rangle\FS}{2},
\end{equation}
where $\mathcal{A} = \int dxdy$. Assuming an infinite and maximally pair-breaking interface with negligible order-parameter suppression, the cost of the midgap states is given by
\begin{equation}
	\label{eq:qc_lowtemp:mgs_energy}
	\frac{\delta\Omega^{\mathrm{MGS}}}{\mathcal{L}\xi_0\NF T_{\mathrm{c}}} = \frac{\pi^2}{2\barvF}\left\langle\left|\mathbf{v}_{\mathrm{F}}\cdot\mathbf{\hat{n}}\right| \left|\Delta(\vkF)\right|\right\rangle\FS \geq 0,
\end{equation}
where $\mathcal{L} = \int dx$, $\barvF \equiv \langle\left|\mathbf{v}_{\mathrm{F}}\right|\rangle\FS$, and $\mathbf{\hat{n}} = \mathbf{\hat{y}}$ is the surface normal. For the circular Fermi surface, this reduces to $\delta\Omega^{\mathrm{MGS}}/(\mathcal{L}\xi_0\NF T_{\mathrm{c}}) = \pi(\sqrt{8}/3)\Delta_0$.
Introducing a small and homogeneous superflow $\mathbf{p}_{\mathrm{s}} = p_{\mathrm{s}} \mathbf{\hat{x}}$ parallel to the interface, the midgap states are Doppler shifted by an energy $\delta_{\epsilon} \propto \mathbf{v}_{\mathrm{F}}\cdot\mathbf{p}_{\mathrm{s}}$, leading to an over-all energy gain
\begin{equation}
	\label{eq:qc_lowtemp:doppler_energy}
	\frac{\delta\Omega^{\mathrm{DS}}}{\mathcal{L}\xi_0\NF T_{\mathrm{c}}} = -\frac{\pi^2}{\barvF}\left\langle\left|\mathbf{v}_{\mathrm{F}}\cdot\mathbf{\hat{n}}\right| \left|\mathbf{v}_{\mathrm{F}}\cdot\mathbf{p}_{\mathrm{s}}\right|\right\rangle\FS \leq 0.
\end{equation}
Again, this can be taken analytically for the circular Fermi surface, yielding the dimensionless expression $\delta\Omega^{\mathrm{DS}}/(\mathcal{L}\xi_0\NF T_{\mathrm{c}}^2) = -\pi\left|\mathbf{p}_{\mathrm{s}}\right|/p_0$, where $p_0 \equiv T_{\mathrm{c}}/\barvF$ is a natural scale for the superflow. Note that the energy gain in Eq.~(\ref{eq:qc_lowtemp:doppler_energy}) is linear in $\left|\mathbf{p}_{\mathrm{s}}\right|$, while the energy cost of bulk superflow scales as $|\mathbf{p}_{\mathrm{s}}|^2$, with the pair-breaking (pb) superflow of the order $\left|\mathbf{p}^{\mathrm{pb}}_{\mathrm{s}}\right| = \Delta_0/\barvF$, see e.g. Refs.~\onlinecite{Holmvall:2019b,Vorontsov:2018}. 
Finally, the angular dependence of the integrands in Eqs.~(\ref{eq:qc_lowtemp:bcs_energy})--(\ref{eq:qc_lowtemp:doppler_energy}) are shown together with the Fermi velocity, $\vF$, in Fig.~\ref{fig:qc_analytic_angle_resolved}. The figure shows that having a large Fermi velocity along the antinodes is conducive to large midgap-state energy costs, as well as large energy gains of Doppler-shifting those states, within the quasiclassical theory of superconductivity.

\bibliography{dwave_grain}
\end{document}